\documentclass[aps,prb,twocolumn,superscriptaddress]{revtex4}
\usepackage{color}
\usepackage{bm}
\usepackage{amsmath}
\usepackage{amssymb}
\usepackage{amsthm}
\usepackage{graphicx}
\usepackage{epsfig}
\usepackage{natbib}

\begin{document}
\title{Elementary excitations in spinor polariton- electron systems}
\author{O. Kyriienko}
\affiliation{Science Institute, University of Iceland, Dunhagi 3,
IS-107, Reykjavik, Iceland}
\author{I. A. Shelykh}
\affiliation{Science Institute, University of Iceland, Dunhagi 3,
IS-107, Reykjavik, Iceland}
\address{International Institute of Physics, Av. Odilon
Gomes de Lima, 1772, Capim Macio, 59078-400, Natal, Brazil}
\date{\today}

\begin{abstract}
We consider effective interactions in a 2D hybrid polariton-
electron system and calculate dispersion of elementary excitations
accounting the spin degree of freedom of the particles. Due to the crucial role played by the exchange term in polariton- electron interactions the dispersions of this system become spin- dependent and show unusual behavior. The coupling of the excitations of the condensate with 2D plasmon can result in appearence of roton minimum in the dispersion and destruction of the condensate for close enough situated quantum wells with electrons and excitons.
\end{abstract}
\maketitle
\section{Introduction}
Hybrid Bose-Fermi systems reveal a number of the quantum collective effects, including superfluidity \cite{Dutta}, polaron effects \cite{Privitera}, Cooper pairing and BCS-BEC crossover \cite{Stoof, Heiselberg, Efimov, Viverit}, formation of supersolid state \cite{Orth} and others. All these phenomena are well studied from theoretical point of view in the domain of cold atoms. However, due to the large masses of the particles the characteristic temperatures of these effects are extremely small (usually in nano- Kelvin range), which makes their experimental observation extremely difficult and rules out any possibilities for their practical implementations.

Recently, it was pointed out that hybrid Bose- Fermi systems posessing intriguing properties can be realized in solid state heterostructures \cite{Laussy,ShelykhRotons,Berman2010}. Due to the much smaller masses of the particles critical temperatures of quantum collective phenomena in this case become orders of magnitude bigger then for cold atoms. From this point of view, the hybrid systems containing exciton (or cavity) polaritons are of special interest, as polaritons have the smallest effective mass among all the particles observed in condensed matter system up to now (about $10^{-4}-10^{-5}$ of the effective mass of the free electron).

Exciton polaritons are quasiparticles that arise from the strong coupling of microcavity photons with quantum-well (QW) excitons \cite{KavokinBook}. Being composite light- matter excitations, cavity polaritons reveal a set of unusual properties. Their extremely small effective mass comes from their photonic component, while the presence of the excitonic component makes possible efficient polariton- polariton and polariton- phonon interactions leading to the thermalization of the polariton ensemble. Combined with bosonic statistics, these leads to extremely high critical temperatures of the polariton BEC (about 20 K in CdTe cavities \cite{Kasprzak} and up to room temperature in GaN cavities \cite{Room-Temp}) and large critical velocities of the polariton superfluid current \cite{Amo, Amo1}.

An important peculiarity of the polariton system is its spin
structure: being formed by bright heavy-hole excitons, the lowest
energy polariton state has two allowed spin projections on the
structure growth axis ($\pm1$), corresponding to the right and left
circular polarisations of the counterpart photons. The states having
other spin projections are split-off in energy and normally can be
neglected while considering polariton dynamics. Thus, from the
formal point of view, the spin structure of cavity polaritons is
similar to the spin structure of electrons, as both are two-level
systems \cite{Shelykh2010}. The possibility to control the spin of
cavity polaritons opens a way to control the polarization of the light
emitted by a cavity, which can be of importance in various
technological implementations including optical information
transfer \cite{Liew}.

Exciton polaritons are electrically neutral and cannot carry electric current. However, they may coexist and interact with free electrons or holes, if these carriers are introduced in the same QW with excitons or in the neighboring QW. When confined together, exciton polaritons and free carriers form Bose-Fermi mixture which is expected to exhibit peculiar optical and electronic properties \cite{Laussy}.

In the present paper we study the effect of spin-dependent interaction of a BEC of microcavity polaritons with a 2DEG on the energy spectrum of elementary excitations in the hybrid system. We consider a quantum microcavity in strong coupling regime, which in addition to the QW with excitonic transition tuned in resonance with photonic cavity mode contains n-doped QW with two dimentional electron gas (2DEG). We restrict our consideration to the regime of the polariton BEC created by non- resonant continous pump, for which quasi-equilibrium treatment can be applied \cite{KasprzakPRL}.

As polaritons are interacting particles, in the absence of free electrons the spectrum of the polariton BEC reveals the Bogoliubov-like renormalization showing linear dispersion near the ground state and parabolic dispersion at larger wave-vectors \cite{Yamamoto}. The situation becomes more complicated if spinor nature of cavity polaritons is accounted for. Due to the predominance of the exchange contribution in the scattering of 2D excitons \cite{Ciuti1998,Combescot2007}, the polariton-polariton interactions become strongly spin- anisotropic, with interaction of the polaritons having same circular polarization (so called 'triplet configuration') being strong and repulsive, and polaritons with opposite circular polarizations ('singlet configuration') - weak and attractive \cite{Renucci}. To minimize its energy polariton condensate forms with linear polarization \cite{Laussy2006} and its dispersion contains two Bogoliubov- like branches with slightly different sound velocities corresponding to the excitations polarized parallel or perpendicular to the polarization of the condensate \cite{Shelykh2006}.

In the present paper we investigate how the dispersions of the elementary excitations of the spinor polariton BEC is modified by coupling with a plasmonic mode of 2DEG. We show that due to the spin dependence of the polariton- electron interactions the spectrum of elementary excitations for two Bogoliubov branches changes in non-symmetrical way. This effect can be viewed as renormalization of the matrix elements of the polariton- polariton interactions in triplet and singlet configurations, which can be tuned by increasing or decreasing BEC-2DEG coupling. Investigation of spectrum as a function of the distance $\Delta$ between QW with excitons and 2DEG shows that at small $\Delta$ the strong polariton- electron interaction leads to the onset of the effective attraction between polaritons leading to the collapse of the condensate phase, while at big $\Delta$ polariton and electron systems become decoupled and standard Bogoliubov and 2D plasmon dispersions are recovered. In the intermediate region, the dispersions of the elementary excitations are qualitatively different from those revealed by uncoupled system. In particular, the difference between sound velocities of the Bogoliubov branches corresponding to different polarizations of the elementary excitations of the condensate is strongly enhanced by polariton- electron coupling. Moreover, the dispersion of the modes can become non- monotonous and reveal roton minimum, similar to those recently predicted for the hybrid system containing spatially separated electrons and indirect excitons \cite{ShelykhRotons}. These radical changes of the dispersions in the presence of 2DEG can strongly affect the real space propagation of the spinor polariton droplets and can be possibly used for spinoptronic applications \cite{PolDevices}. The analysis of these perspectives, however, lies beyond the scope of the present paper.

The article is organized as follows. In Section II we discuss effective interactions in spinor polariton-electron mixtures. In Section III we obtain the expressions of elementary excitations which are analyzed in detail in Section IV. Conclusions summarize the results of the work.

\section{Effective interactions in a hybrid Bose- Fermi system}
\begin{figure}
\includegraphics[width=1.0\linewidth]{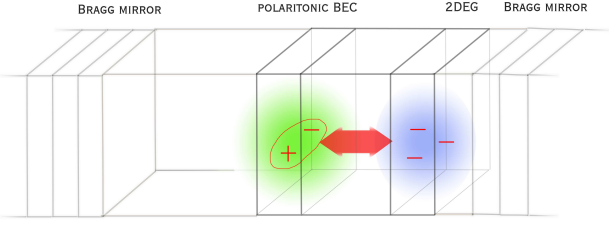}
\caption{Illustration of microcavity structure with polaritonic condensate (green) and 2D electron gas (blue). The photons are confined between dielectric Bragg mirrors and strongly interact with excitonic transition in the QW placed at the place where the electric field of the cavity mode have maximum. Strong coupling between photons and excitons leads to the formation of cavity polaritons. Free electrons are situated in the QW parallel to the QW with excitons and interact with them, which leads to the coupling of electronic and polaritonic systems.}
\label{Fig1}
\end{figure}

\begin{figure}
\includegraphics[width=1.0\linewidth]{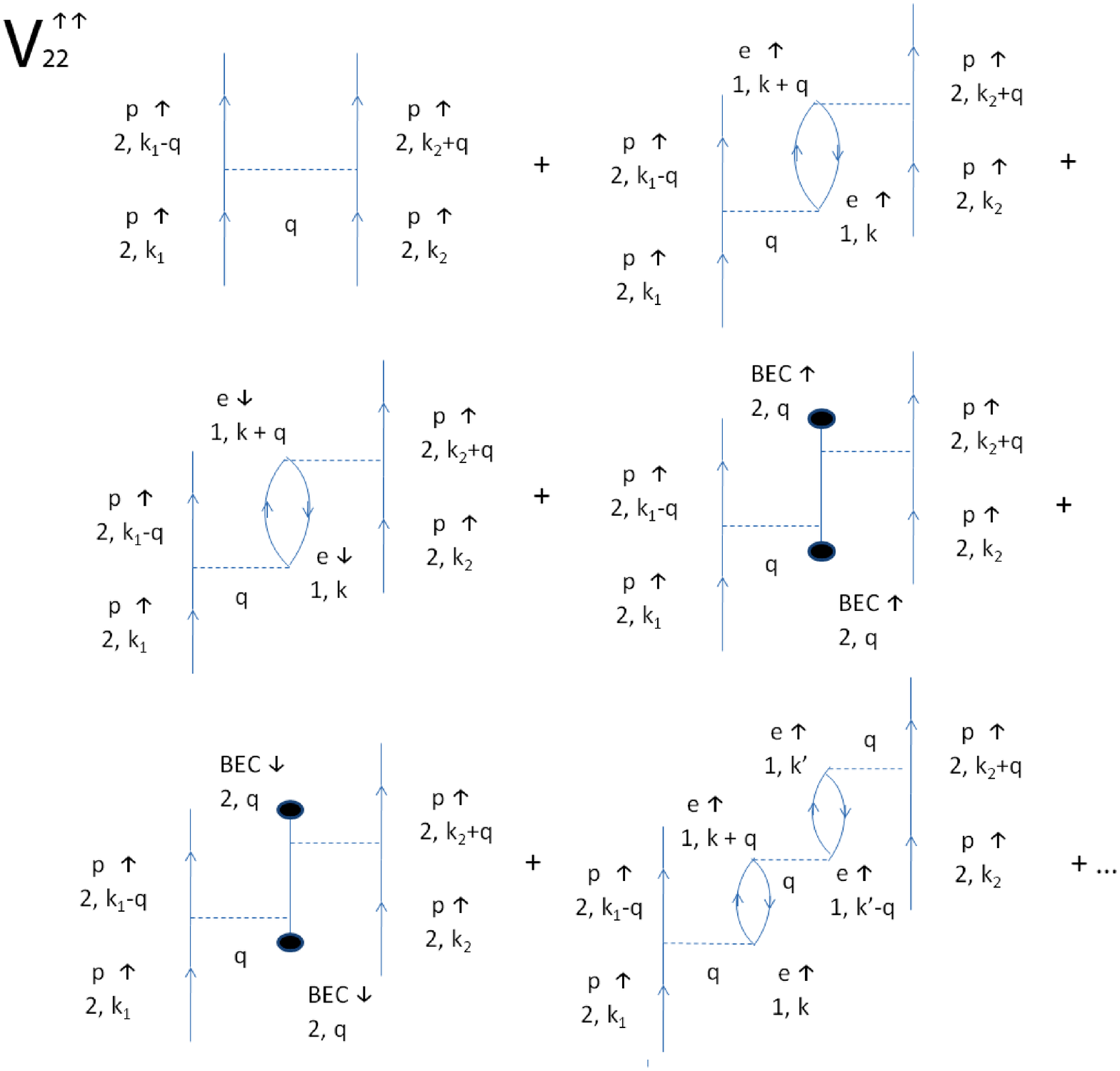}
\caption{Feynman diagrams for the matrix element of the effective polariton-polariton interaction. Index 1 corresponds to electrons, index 2 to the exciton-polaritons. Arrow shows the spin projection. Black circles correspond to the polariton BEC.}
\label{Fig2}
\end{figure}

In this section we present the method of calculating the spin-dependent effective interactions in multilayered systems and apply it for the determination of spectrum of the elementary excitations.
Consider a system of two parallel semiconductor quantum wells (QWs), one of which contains a free electron gas, and other containing a BEC of excitons embedded into a photonic cavity (Fig.\ref{Fig1}).
The effective interaction between electrons and polaritons accounting their degree of freedom can be calculated analogicaly to spinless case\cite{ShelykhRotons}. The effective matrix element of the interaction can be represented as a sum of the infinite series of the diagrams. In the random phase approximation (RPA) and in the assumption that polariton condensate is weakly depleted, only certain types of the diagrams should be retained. For the electron system, these are diagrams corresponding to the excitations of virtual electron- hole pairs in a Fermi sea (so called "polarization bubbles"). For the polariton system these are diagrams corresponding to the interaction of the particle with a condensate accompanied by creation its excitation. The resulting set is shown at on Fig.\ref{Fig2}.

Evaluation of Feynman diagrams by using standard rules \cite{Zagoskin} gives the 4$\times 4$ effective interaction matrix accounting for the two possible projections for the electrons ($\pm\frac{1}{2}$) and polaritons ($\pm 1$) and can be represented as
\begin{equation}
\mathbf{V}^{eff}=\mathbf{V}\cdot \left( 1-\Pi \mathbf{V}\right) ^{-1}
\label{Veff}
\end{equation}
where $\mathbf{V}^{eff},\mathbf{V}$ are $4\times 4$ matrices
of bare spin- dependent interactions and $\Pi $ is diagonal polarization matrix with the diagonal elements of electronic ($\Pi_{1}$) and excitonic ($\Pi_{2}$) polarizations.
\begin{equation}
\mathbf{V} =
\left( \begin{array}{cccc}
V_{11} & V_{11} & V_{12}^{\uparrow\uparrow} & V_{12}^{\uparrow\downarrow} \\
V_{11} & V_{11} & V_{12}^{\uparrow\downarrow} & V_{12}^{\uparrow\uparrow} \\
V_{12}^{\uparrow\uparrow} & V_{12}^{\uparrow\downarrow} & V_{22}^{\uparrow\uparrow} & V_{22}^{\uparrow\downarrow} \\
V_{12}^{\uparrow\downarrow} & V_{12}^{\uparrow\uparrow} & V_{22}^{\uparrow \downarrow} & V_{22}^{\uparrow\uparrow} \\
\end{array} \right)
\end{equation}
\begin{equation}
\mathbf{\Pi} =
\left( \begin{array}{cccc}
\Pi_{1} & 0 & 0 & 0 \\
0 & \Pi_{1} & 0 & 0 \\
0 & 0 & \Pi_{2} & 0 \\
0 & 0 & 0 & \Pi_{2} \\
\end{array} \right)
\end{equation}
where $V_{11}, V_{22}, V_{12}$ correspond to electron-electron, polariton-polariton and electron-polariton bare matrix elements, respectively. $\Pi_{1}$ and $\Pi_{2}$ are polarization operators for electronic and polaritonic systems. We assumed that there is no external magnetic field applied to the system and thus the concentration of spin-up and spin down components are the same.

The dispersion of the elementary excitations is given by condition
\begin{equation}
\mathbf{det} \left( 1-\Pi \mathbf{V}\right)=0
\label{det}
\end{equation}
In the case when bosonic and fermionic subsystems are decoupled, the solution of Eq.\ref{det} corresponds to the two spin-dependent Bogoliubov branches of the dispersion of the polariton BEC, and spin- independent plasmonic mode. In a coupled system the interaction between polaritons and electrons leads to the formation of the hybrid plasmon- polariton modes, as it is discussed below.

Let us now define all matrix elements used in further calculations. The electron-electron interaction in two dimensional electron gas is described by 2D Coulomb potential,
\begin{equation}
V_{11}(q)=\frac{e^{2}}{2\epsilon _{0}\epsilon A}\cdot \frac{1}{q},
\label{V11}
\end{equation}
where $A$ being the sample area and $\epsilon $ being a dielectric constant of the media and $q$ is transferred momentum.

The matrix elements of the polariton-polariton interaction $V_{22}$ is spin-dependent due to strong exchange contribution \cite{Ciuti1998}. They depend on the geometry of the cavity and detuning between polariton and exciton modes\cite{Ouerdane,Ostatnicky,Masha}.  Usually the matrix element for polariton interaction in anti-parallel spin configuration has opposite sign and is about one order of magnitude smaller than for parallel spins \cite{Renucci,Shelykh2010}. The simple estimation of the matrix element of triplet polariton-polariton interaction which we use in a present paper can be obtain as \cite{Tassone}
\begin{equation}
V_{22}\approx \frac{6E_{B}X^{2}a_{B}^{2}}{A},
\label{V22}
\end{equation}
where $E_{B}$, $a_{B}$ and $X$ being the exciton binding energy, Bohr radius and the exciton Hopfield coeficient, respectively. The matrix element of the interaction is singlet configuration is taken to be negative and 10 times smaller then for the triplet configuration, in agreement with the results of Ref.\onlinecite{Renucci}.

The coupling between bosonic and fermionic systems is determined by matrix elements of the electron- polariton interaction. The interaction matrix element consists of the direct and exchange parts:
\begin{widetext}
\begin{equation}
V_{dir}(\mathbf{q})=\int \Psi_{X}^{\star}(\mathbf{Q},\mathbf{r_{e}},\mathbf{r_{h}})
\Psi_{e}^{\star}(\mathbf{k},\mathbf{r_{c}})
V(\mathbf{r_{c}},\mathbf{r_{e}},\mathbf{r_{h}})
\Psi_{X}(\mathbf{Q},\mathbf{r_{e}},\mathbf{r_{h}})
\Psi_{e}(\mathbf{k},\mathbf{r_{c}})d\mathbf{r_{c}}d\mathbf{r_{e}}d\mathbf{r_{h}},
\label{Vdir}
\end{equation}

\begin{equation}
V_{exc}(\mathbf{q})=\int \Psi_{X}^{\star}(\mathbf{Q},\mathbf{r_{e}},\mathbf{r_{h}})
\Psi_{e}^{\star}(\mathbf{k},\mathbf{r_{c}})
V(\mathbf{r_{c}},\mathbf{r_{e}},\mathbf{r_{h}})
\Psi_{X}(\mathbf{Q},\mathbf{r_{c}},\mathbf{r_{h}})
\Psi_{e}(\mathbf{k},\mathbf{r_{e}})d\mathbf{r_{c}}d\mathbf{r_{e}}d\mathbf{r_{h}},
\label{Vexc}
\end{equation}
\end{widetext}
where $\mathbf{Q}$ and $\mathbf{k}$ are exciton and free electron momentum, and $\mathbf{r_{e}}$, $\mathbf{r_{h}}$, $\mathbf{r_{c}}$ are radius vectors of bounded electron, bounded hole and free electron, respectively. The matrix elements $V_{12}^{\uparrow\downarrow},V_{12}^{\uparrow\uparrow}$ can be expressed as
\begin{eqnarray}
V_{12}^{\uparrow\downarrow}=V_{dir}+V_{exc},\\
V_{12}^{\uparrow\uparrow}=V_{dir}
\end{eqnarray}

\begin{figure}
\includegraphics[width=1.0\linewidth]{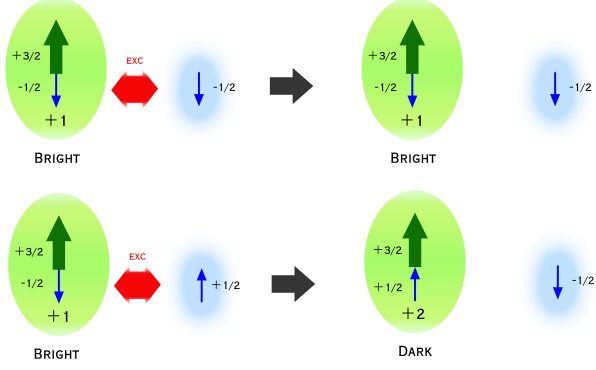}
\caption{The interaction of the exciton and free carrier accompanied by exchange of the electrons. In the configuration where the spins of the exchanged electrons are antiparallel the process leads to the transition of the exciton towards dark state, which becomes unefficient due to the huge Rabi splitting between polariton and dark exciton modes.}
\label{Fig3}
\end{figure}

The reason why the exchange term enters only into matrix element $V_{12}^{\uparrow\downarrow}$ is illustrated at Fig.\ref{Fig3}. As one sees, the exchange of electrons between exciton and electron for configurations $(\pm1\mp1/2)$ does not change their spin states, while for configurations $(\pm1\pm1/2)$ it leads to the transition of the exciton into dark state. In microcavity this process is blocked because dark states are uncoupled to a cavity mode and thus splitted in energy from polariton states by several meV.

For the case of separated QWs it is convinient to use 2D in-plane Fourier transforms of wave functions of the electron and exciton to separate the free in-plane motion from the confinement in z direction. Then, exciton wave function is
\begin{equation}
\phi_{\mathbf{k}}(z_{e},z_{h})=\sqrt{\frac{2}{\pi a_{B}^{2}}}[1+(a_{B}k)^2]^{-3/2}\chi_{e}(z_{e})\chi_{h}(z_{h}),
\label{wavefuncX}
\end{equation}
where $\chi_{e}$ and $\chi_{h}$ are confinement functions in $z$ direction for electron and hole in exciton, respectively. The free electron confinement function $\chi_{c}$ describes the eletron gas separated by distance $\Delta$ from BEC. These functions in the region between wells were supposed to decay exponentially, $\chi(z)=e^{-\frac{|z|}{L_{0}}}$, with $L_{0}$ being characteristic decay length determined by the geometry of the structure.

Following the scheme, proposed in the Ref. \onlinecite{Ramon2003} we can rewrite direct term (\ref{Vdir}) as
\begin{equation}
V_{dir}(q)=\frac{4e^{2}}{\epsilon\epsilon_{0}AqL^{3}}[g(a_{B}\beta_{e}q/2)-g(a_{B}\beta_{h}q/2)]\mathit{I_{z}^{dir}}
\label{Vdir2}
\end{equation}
where $g(q)=[1+(a_{B}q)^{2}]^{-3/2}$, $\beta_{e,h}=m_{e,h}/M_{X}$, $\Delta$ is equal to distance between centers of QWs, and
\begin{eqnarray}
\mathit{I_{z}^{dir}}=\int_{-\infty}^{\infty}dz_{h}dz_{e}dz_{c}e^{-q\Delta}e^{\frac{-2|z_{h}|}{L_{0}}}e^{\frac{-2|z_{e}|}{L_{0}}}e^{\frac{-2|z_{c}-\Delta|}{L_{0}}}=\\
\nonumber=e^{-q\Delta}L_{0}^{3}
\label{intZdir}
\end{eqnarray}

The exchange interaction matrix element can be derived in similar way and recast as
\begin{equation}
V_{exc}(q)=\frac{2e^{2}}{\pi^{2}\epsilon\epsilon_{0}AL^{2}}\mathit{I(q)I_{z}^{exc}}
\label{Vexc3}
\end{equation}
where
\begin{eqnarray}
\mathit{I_{z}^{exc}}=\int_{-\infty}^{\infty}dz_{h}dz_{e}dz_{c}e^{\frac{-2|z_{h}|}{L_{0}}}e^{\frac{-|z_{e}|-|z_{e}-\Delta|}{L_{0}}}e^{\frac{-|z_{c}-\Delta|-|z_{c}|}{L_{0}}}=\\
\nonumber=L_{0}(L_{0}+\Delta)^{2}e^{-\frac{2\Delta}{L_{0}}}
\label{intZexc}
\end{eqnarray}
and $\textit{I(q)}$ is dimensionless integral studied in details in Ref.\onlinecite{Ramon2003} which can not be computed analytically. Differently from the direct interaction term, the exchange term contains the factor $e^{-\frac{2\Delta}{L_{0}}}$ coming from the overlap integral between wavefunctions of the electron forming the exciton and free electron from 2DEG and decays exponentially with increase of the distance between QWs even at $q=0$.

\begin{figure}
\includegraphics[width=1.0\linewidth]{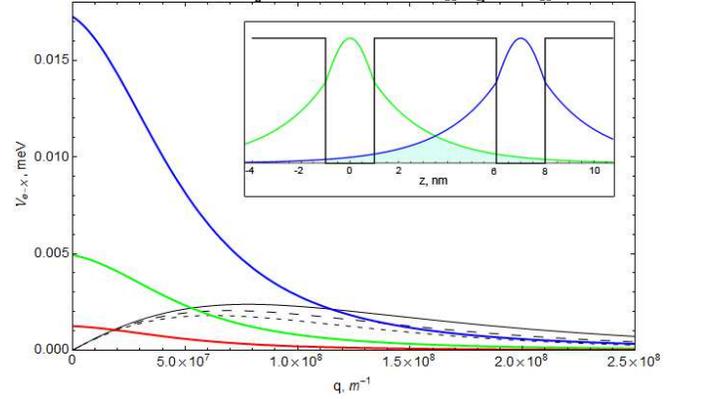}
\caption{Momentum dependence of electron-polariton matrix element. Black and blue thick lines correspond to direct and exchange matrix elements for separation of wells $\Delta=4nm$, respectively. Green and red line show exchange interaction for wells separation $\Delta = 6 nm$ and $8 nm$, respectively.
Inset: The confinement in the hybrid system of separated quantum wells with width $L=2 nm$. Characteristic decreasing length was estimated as $2.3 nm$. Filled area under the curves demonstrates the wave function overlap.}
\label{Fig4}
\end{figure}

The dependencies of the direct and exchange matrix elements on transferred momentum $q$ are shown at Fig.\ref{Fig4}. One sees that for $q=0$ the direct term goes to zero, while exchange term remains finite. In general, for small enough momenta the exchange term dominates over the direct one if separation between the wells containing excitons and 2DEG is below 15-20 nanometers.\\ \\
\section{Dispersion of the elementary excitations}
The dispersions of the elementary excitations are determined by the polarization operators for the condensate and 2D electron gas. For the polaritonic BEC the polarization is defined as \cite{Zagoskin}
\begin{equation}
\Pi _{2}(\mathbf{q},\omega )=N_{0}G_{0}^{ex}(\mathbf{q},\omega )=\frac{
2N_{0}E_{\mathbf{q}}^{pol}}{(\hbar \omega )^{2}-(E_{\mathbf{q}}^{pol})^{2}}
\label{Pi2}
\end{equation}
where $N_{0}$ is the occupation number of the condensate and $E_{\mathbf{q}
}^{pol}$ is dispersion of the bare excitons (parabolic, in the effective mass approximation).

In general, polarization operator of electron system can be written using Lindhard formula \cite{Lindhard}
\begin{equation}
\Pi _{1}(\mathbf{q},\omega )=\sum_{\mathbf{k}}\frac{f_{\mathbf{k-q}}-f_{%
\mathbf{k}}}{\hbar (\omega +i\delta) +E_{\mathbf{k-q}}^{el}-E_{\mathbf{k}%
}^{el}},
\label{Pi1}
\end{equation}
where  $E_{\mathbf{q}}^{ex}$ and $E_{\mathbf{q}}^{el}$ are dispersions of the bare excitons and electrons, respectively, and $f_{\mathbf{k,}}$, $f_{\mathbf{k-q}}$ are Fermi distributions.

First, one can analyse the simplest case, when one uses the spproximation of the static screening. In this case the polarization operatorator of 2DEG at zero temperature reads \cite{Koch}
\begin{equation}
\Pi _{1}^{0}\approx - A\frac{m_{el}}{\pi \hbar ^{2}}
\label{Pi1stat}
\end{equation}
where $m_{el}$ is the electron effective mass.
In this case after straigthforward algebra one can obtain two solutions of Eq.\ref{det}:
\begin{widetext}
\begin{equation}
\hbar \omega _{U}(q)=\sqrt{E_{pol}^{2}(q)+N_{0}E^{pol}(q)\left[(V_{22}^{\uparrow\uparrow}-V_{22}^{\uparrow\downarrow})+
\Pi_{1}(V_{12}^{\uparrow\uparrow}(q)-V_{12}^{\uparrow\downarrow}(q))^{2}\right]}
\label{upper}
\end{equation}
\begin{equation}
\hbar \omega _{L}(q)=\sqrt{E_{pol}^{2}(q)+N_{0}E^{pol}(q)\left[(V_{22}^{\uparrow\uparrow}+V_{22}^{\uparrow\downarrow})+
\frac{\Pi_{1}(V_{12}^{\uparrow\uparrow}(q)+V_{12}^{\uparrow\downarrow}(q))^{2}}{1-2\Pi_{1}V_{11}(q)}\right]}
\label{lower}
\end{equation}
\end{widetext}
We considered the case of the linear polarized condensate for which the concentrations of both circular polarized components are equal.

Naturally, for the decoupled electronic and excitonic systems these equations transform into expressions for spin- dependent dispersions of linear polarized polariton BEC obtained in Ref.\onlinecite{Shelykh2006}. The upper branch corresponds to the excitations polarized perpendicular to the polarization of the condensate, and the lower branch - to the excitations polarized parallel to it. Note, that in our calculation we assume that $V_{22}^{\uparrow\downarrow}<0$.

One can make the following remarks. First, for both at small $q$ the dependence  is linear, $\omega_{L,U}(q)=v_{L,U}q$, which corresponds to the superfluid nature of the system. Second, the interaction with electrons changes polariton dispersion branches in non-symetrical way. Due to the crucial role played by electron exchange these interactions are spin- dependent, $V_{12}^{\uparrow\uparrow} \neq V_{12}^{\uparrow\downarrow},|V_{22}^{\uparrow\uparrow}| \neq |V_{22}^{\uparrow\downarrow}|$. Indeed, for the co- polarized lower branch (\ref{lower}) there is no difference from the dispersion of the polariton BEC uncoupled from electronic system and corresponding sound velocity remains the same, $v_L=\sqrt{N_0(V_{22}^{\uparrow\uparrow}+V_{22}^{\uparrow\downarrow})/2m}$. This is because of the long range character of the electron-electron interactions $\Pi_{1}(V_{12}^{\uparrow\uparrow}(0)+V_{12}^{\uparrow\downarrow}(0))^{2}/(1-2\Pi_{1}V_{11}(0))\sim q$, when $q\rightarrow0$. On the other hand, for the upper branch polariton electron interactions renormalize the sound velocity, which becomes $v_U=\sqrt{[N_0(V_{22}^{\uparrow\uparrow}-V_{22}^{\uparrow\downarrow})+\Pi_{1}(V_{12}^{\uparrow\uparrow}(0)-V_{12}^{\uparrow\downarrow}(0))^{2}]/2m}$.

The above consideration neglected the dynamic effects in the polarizability of the electronic system. The latter, however, can lead to the appearance of the plasmonic mode in 2DEG, which, differently from the plasmonic mode in 3D case is gapless \cite{Koch,Stern}, $\omega_{pl}\sim\sqrt{q}$. One can thus expect that this mode can be in principle strongly coupled with Bogoliubov modes of the condensate and hybrid plasmon-bogolon elementary excitations appear. To find there dispersion, in Eqs. \ref{upper},\ref{lower} $\Pi_1$ should be considered as a function of $\omega$. These expressions thus represent some transcedent equations, one of which have a single solution (Eq.\ref{upper}), while another one - two solutions (Eq.\ref{lower}).

To account for the dynamic effects in polarizability of 2DEG we used the results of the Ref.\onlinecite{Stern}. According to this work, one has:
 \begin{equation}
\Pi _{1}=\Pi_{1}^{0}(1-C_{-}[(z-u)^{2}-1]^{1/2}-C_{+}[(z+u)^{2}-1]^{1/2})
\label{Pi1dyn}
\end{equation}
where $z=q/2k_{F}$ and $u=\omega/q\mathit{v}_{F}$ are dimensionless quantities. Coefficients $C_{\pm}=Sign(z \pm u)$ for $|z \pm u|>1$ and $C_{\pm}=0$ for $|z \pm u|<1$.
Using this polarization operator one can easily obtain the spectrum of collective excitations in 2D electron gas with square-root dispersion (Fig.\ref{Fig5}, inset). For the coupled polariton- electron system the dispersions of the elementary excitations accounting for dynamics effects can be found only numerically and are presented in the following section.

\section{Results and discussion}
\begin{figure}
\includegraphics[width=1.0\linewidth]{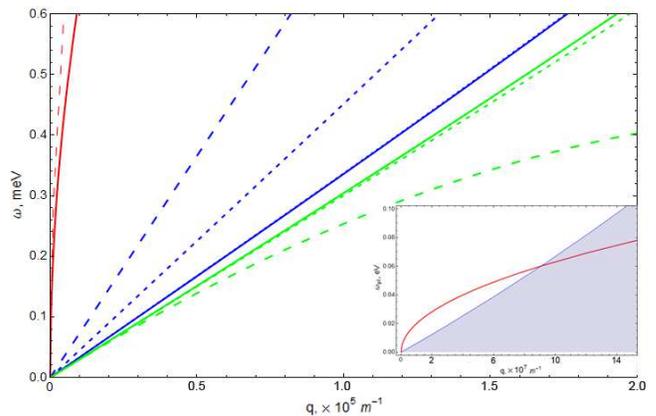}
\caption{Linear dispersion of the elementary excitations of the condensate in region of small momentum. $\Delta = 4, 6 nm$ (large and medium dashed lines, respectively). Solid lines represent bare upper (blue) and lower (green) BEC dispersion branches. Red solid and dashed lines correspond to initial and modified plasmon modes.
Inset: Dispersion of plasmon mode in singe quantum well with 2DEG (red line). Blue filling indicates electron-hole excitation continium.}
\label{Fig5}
\end{figure}

In the present work we consider GaAs cavity at $T=0$ with the concentration of the polariton condensate $n_{0}=10^{11}cm^{-2}$. The plot on the Fig.\ref{Fig5} shows the dispersions of the three hybrid branches for different separation $\Delta$ (dashed lines) compared to the dispersions of the condensate in the absence of the electrons (blue and green solid lines). One sees that the presence of the electron gas changes the slope of the curve and thus the value of superfluid velocity.

\begin{figure}
\includegraphics[width=1.0\linewidth]{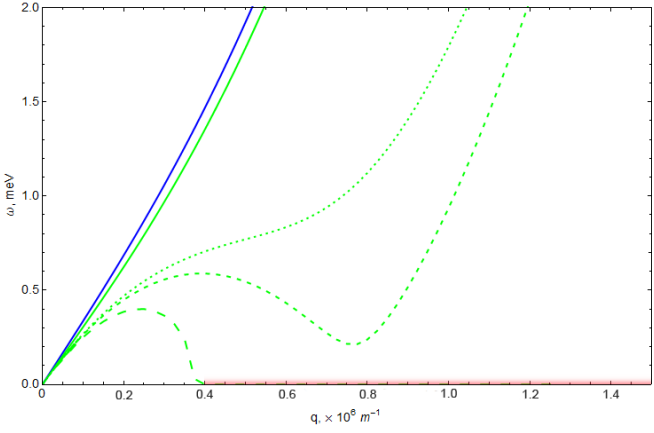}
\caption{Dispersion of the elementary excitations of the condensate, plotted for standard parameters of GaAs/AlAs heterostructure. Solid blue and green lines correspond unperturbed upper and lower dispersion branches. Dashed lines (large-to-small dashing) show the process of condensate braking via formation of the roton-like minimum (for $\Delta=4.2nm$ to $4nm$).}
\label{Fig6}
\end{figure}

Comparing to the static case, there are two important differencies. First, third branch corresponding to modified plasmon dispersion appears (red dashed line). Second, under certain conditions, the lower branch can have roton minimum this the spectrum as it is shown at Fig.\ref{Fig6}. With decrease of the separation $\Delta$ the roton minimum becomes deeper, and below some critical separation can pass below zero, which indicates the instability of the condensate due to the onset of the effective attraction between the polaritons mediated by virtual excitations of the electron system, analogicaly to those predicted for indirect excitonic system \cite{ShelykhRotons}. When $\Delta$ becomes large ($>10nm$), the instability and roton minimum disappear (small dashed line) and the dispersion of the elementary excitations recovers the dispersion corresponding to the case when no electrons are present.
\\

\begin{figure}
\includegraphics[width=1.0\linewidth]{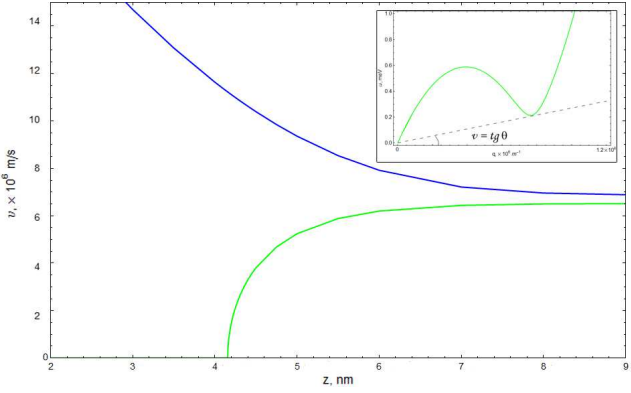}
\caption{Polariton condensate velocities as a function of separation distance of quantum wells. Blue solid line shows the corresponding velocity of upper branch, while green curve shows dependence of the lower branch velocity.}
\label{Fig7}
\end{figure}

The dependence of superfluid velocity on distance between 2DEG and excitons is shown at Fig.\ref{Fig7}. For small distances condensate is instable due to the presence of the deep roton minimum. After the stability is regained, there is a region of two superfluidic phases with different sound velocities.  Finally,  at very big $\Delta$ the velocities reach constant values when BEC and 2DEG are uncoupled. Note, that for the lower branch the superfluid velocity is determined not by the slope of the dispersion at $k=0$ but by a roton minimum (if it is present).

\begin{figure}
\includegraphics[width=1.0\linewidth]{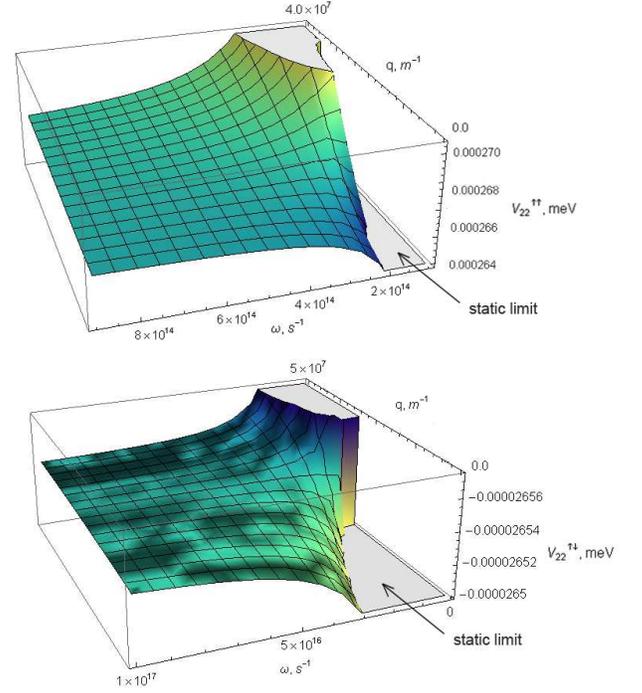}
\caption{The effective interaction constants $\tilde{V}_{22}^{\uparrow\uparrow,\uparrow\downarrow}$ as a function of momentum and frequency. While for static case the constants are strongy modified by polariton- electron interaction (gray region, $V_{22}^{\uparrow\uparrow,\uparrow\downarrow} \approx |0.05| meV$), for high frequency region retardation effects play role and effective constants correspond to initial.}
\label{Fig8}
\end{figure}

The difference between the dispersions of hybrid polariton- electron system and polariton condensate can be described in terms of renormalization of the polariton- polariton interaction constants in singlet and triplet configurations. The renormalized constants can be found as
\begin{eqnarray}
\tilde{V}_{22}^{\uparrow\uparrow}=V_{22}^{\uparrow\uparrow}+\frac{\Pi_1(q,\omega)}{2}\left[(V_{12}^{\uparrow\uparrow}(q)-V_{12}^{\uparrow\downarrow}(q))^{2}\right.+\\
\nonumber\left.+\frac{(V_{12}^{\uparrow\uparrow}(q)+V_{12}^{\uparrow\downarrow}(q))^{2}}{1-2\Pi_{1}V_{11}(q)}\right],\\
\tilde{V}_{22}^{\uparrow\downarrow}=V_{22}^{\uparrow\downarrow}-\frac{\Pi_1(q,\omega)}{2}\left[(V_{12}^{\uparrow\uparrow}(q)-V_{12}^{\uparrow\downarrow}(q))^{2}\right.+\\
\nonumber\left.-\frac{(V_{12}^{\uparrow\uparrow}(q)+V_{12}^{\uparrow\downarrow}(q))^{2}}{1-2\Pi_{1}V_{11}(q)}\right]
\end{eqnarray}
As functions of $q$ and $\omega$, they are plotted on Fig.\ref{Fig8}.  Due to the peculiarities of dynamical polarization in the two dimensional electron gas connected to retardation effects, the sign of efective constants changes in certain frequency range. For example, they can have opposite sign comparing to non- renormalized values in the region of small $q$ and $\omega$. For large frequencies and momemta ($q>10^{9} m^{-1}$) they recover initial values $V_{22}^{\uparrow\uparrow,\uparrow\downarrow}$ (Fig.\ref{Fig8}).

\begin{figure}
\includegraphics[width=1.0\linewidth]{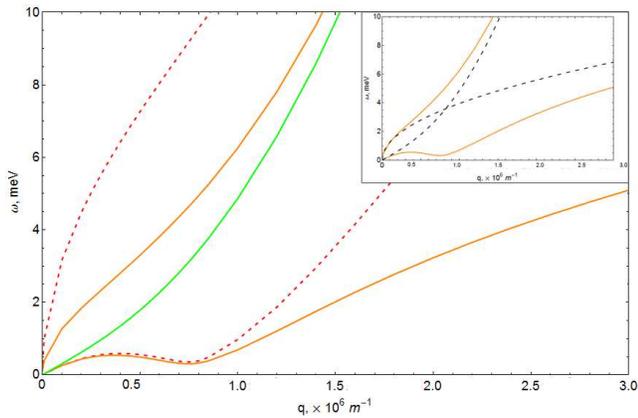}
\caption{The spectrum of elementary excitations for various concentrations of electron gas. For high electron concetrations ($n_{el}=5 \times 10^{12} cm^{-2}$) in the interesting region of $q$ the distinct plasmon and polariton branches exist (red dashed lines), while for small concentration of electron gas ($n_{el}=10^{9} cm^{-2}$) bogolon branch corresponds to those of the uncoupled system (solid green line). In the range of medium concentrations ($n_{el}=8 \times 10^{11} cm^{-2}$) the plasmonic and polaritonic modes are strongly coupled (orange solid lines) and reveal anticrossing behaviour described at the inset of the figure. Inset shows characteristic mode anticrossing for plasmon and polariton dispersion branches (black dashed lines correspond to spectrum of excitations in uncoupled systems).}
\label{Fig9}
\end{figure}

We also analyzed how the spectrum of the elementary excitations changes with variation of the concentration of 2DEG (Fig.\ref{Fig9}). We show only the solutions of Eq.\ref{lower} corersponding to strongly coupled lower Bogoliubov and plasmon modes.
For concentrations about $n_{el}=5 \times 10^{12} cm^{-2}$ plasmon and polariton branches exist. If we put electron concentration very small (about $n_{el}=10^{9} cm^{-2}$), dispersion corresponds to the polariton BEC branch close to initial, while plasmon branch is almost invisible. For the intermediate case, there appears a strong coupling between plasmonic and one of the Bogoliubov modes, characterized by their anticrossing \cite{Tamm}.

We propose several possibilities of tuning the splitting between dispersion branches. First, it can be done by controlling of polarization of cavity exciton polaritons by using elliptically polarized continous pumping or by application of strong magnetic fields \cite{RuboPLA}.

 \begin{figure}
\includegraphics[width=1.0\linewidth]{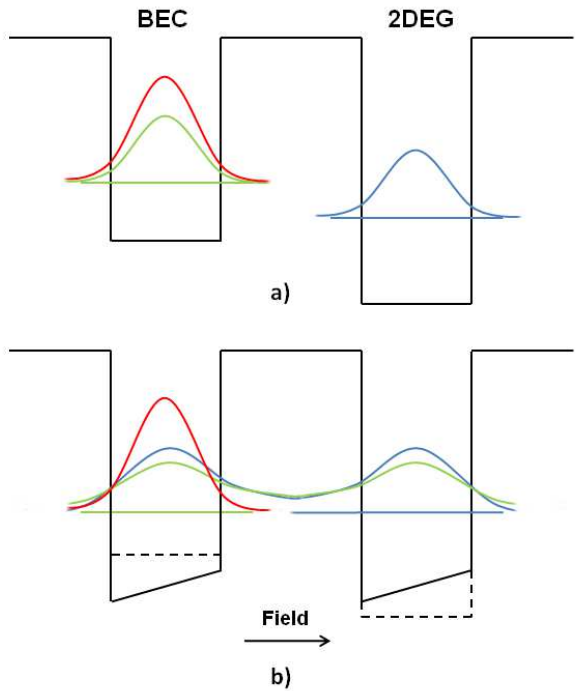}
\caption{The double-barrier structure for obtaining the strongly coupled exciton- electron system. a) Unbiased system, when there is a difference between the energy levels in right and left QWs and electron (blue) and exciton (green, red) wavefunctions overlap only weakly. b) Applying of the bias voltage $V_g$ increases the electron levels of the right quantum well, producing a tunnelling resonance when it matches the left quantum well electron energy. At resonance, the electron wavefunctions will be spread between both wells hereas the the left quantum well hole remains strongly confined. As a result, the exciton aquires significant dipole moment and overplap between wavefunctions of free electron and exciton increases, thus increasing exchange interaction between them.}
\label{Fig10}
\end{figure}

Second, one can use double-barrier structures to control an overlap between wavefunctions of electrons in 2DEG and electrons in excitonic system as it is shown at Fig.\ref{Fig10}. Locating the electron gas in right QW and excitons in left QW and making favorable conditions for free electrons to tunnel through the barrier by changing the gate voltage, it is possible to control electron-polariton exchange interaction by controling the overlap between the wavefunctions as it is shown at Fig.\ref{Fig10}. Recently, such structures were proposed to create polaritons with high $Z$ projections of the dipole moment\cite{Christmann,Oriented}.

 According to the results of this paper, this will change the spectrum of elementary excitations. The latter strongly affects the real space propagation of the polariton droplets \cite{Shelykh2006}. As for the hybrid system the difference between the dispersions of the excitations co and cross polarized to the polarization of the condensate is strongly enhanced compare to isolated polariton BEC, one can expect that a variety of spin- related phenomena will emerge in real space dynamics of polariton- electron mixtures. The detailed investigation of this issue lies, however, beyond the scopes of the present paper.

\section{Conclusions} In conclusion, we analyzed the effective interactions and elementary excitations in hybrid polariton- electron system accounting for the spin degree of freedom of the particles. We have shown that the exchange interaction plays a crucial role, and can lead to nontrivial phenomena, including the instability of the condensate for small electron- exciton separation and appearence of the hybrid plasmon- bogolon modes revealing roton minimum. Besides, it was shown that splitting between elementary excitations with orthogonal linear polarizations is strongly enhanced as compare to isolated polariton system, which can be used for the purposes of spinoptronics.

We thank Alexey Kavokin for discussions on the topic. This work was supported by Rannis "Center of Excellence in Polaritonics", FP7 IRSES projects "POLAPHEN" and "SPINMET" and joint RFBR-CNRS project.


\begin{thebibliography}{99}

\bibitem{Dutta} O. Dutta and M. Lewenstein, Phys. Rev. A \textbf{81}, 063608 (2010).

\bibitem{Privitera} A. Privitera and W. Hofstetter, Phys. Rev. A \textbf{82}, 063614 (2010).

\bibitem{Stoof} M.J. Bijlsma, B.A. Heringa, and H.T.C. Stoof, Phys.
Rev. A \textbf{61}, 053601 (2000).

\bibitem{Heiselberg} H. Heiselberg, C. J. Pethick, H. Smith, and L. Viverit, Phys. Rev. Lett. \textbf{85}, 2418 (2000).

\bibitem{Efimov} D. V. Efremov and L. Viverit, Phys. Rev. B \textbf{65}, 134519 (2002).

\bibitem{Viverit} L. Viverit, Phys. Rev. A \textbf{66}, 023605 (2002͒).

\bibitem{Orth} P.P. Orth, D.L. Bergman, and K. Le Hur, Phys. Rev. A \textbf{80}, 023624 (2009).

\bibitem{Laussy} F. P. Laussy, A. V. Kavokin and I. A. Shelykh, Phys. Rev. Lett. \textbf{104}, 106402 (2010).

\bibitem{ShelykhRotons} I. A. Shelykh, T. Taylor and A. V. Kavokin, Phys. Rev. Lett. \textbf{105}, 140402 (2010).

\bibitem{Berman2010} O.L. Berman, R.Ya. Kezerashvili, and Y.E. Lozovik, Phys. Rev. B \textbf{82}, 125307 (2010).

\bibitem{KavokinBook} A. Kavokin, J.J. Baumberg, G. Malpuech, and F.P. Laussy, Microcavities (Oxford University Press, Oxford, 2007).

\bibitem{Kasprzak} J. Kasprzak, M. Richard, S. Kundemann, A. Baas, P. Jeambrun, J.M.J. Keeling, F.M. Marchetti, M.H. Szymanska, R. Andre, J.L. Staehli, V. Savona, P.B. Littlewood, B. Deveaud, and Le Si Dang, Nature (London) \textbf{443}, 409 (2006).

\bibitem{Room-Temp} S. Christopoulos, G. Baldassarri Hoger von Hoegerstal,
A. Grundy, P.G. Lagoudakis, A.V. Kavokin, J.J. Baumberg,
G. Christmann, R. Butte, E. Feltin, J.-F. Carlin, and N. Grandjean, Phys. Rev. Lett. \textbf{98}, 126405 (2007).

\bibitem{Amo} A. Amo, D. Sanvitto, F. P. Laussy, D. Ballarini, E. del Valle, M. D. Martin, A. Lemaître, J. Bloch, D. N. Krizhanovskii, M. S. Skolnick, C. Tejedor and L. Vina, Nature \textbf{457}, 291 (2009).

\bibitem{Amo1} A. Amo, J. Lefrere, S. Pigeon, C. Adrados, C. Ciuti, I. Carusotto, R. Houdre, E. Giacobino, A. Bramati, Nature Physics \textbf{5}, 805 (2009).

\bibitem{KasprzakPRL} J. Kasprzak, D. D. Solnyshkov, R. Andre, Le Si Dang, and G. Malpuech, Phys. Rev. Lett. \textbf{101}, 146404 (2008).

\bibitem{Shelykh2010} I. A. Shelykh, A. V. Kavokin, Yuri G. Rubo, T. C. H. Liew and G. Malpuech, Semicond. Sci. Technol. \textbf{25}, 013001 (2010).

\bibitem{Liew} T. C. H. Liew, A. V. Kavokin, and I. A. Shelykh, Phys. Rev. Lett. \textbf{101}, 016402 (2008).

\bibitem{Yamamoto} S. Utsunomiya, L. Tian, G. Roumpos, C. W. Lai, N. Kumada, T. Fujisawa, M. Kuwata-Gonokami, A. Loffler, S. Hofling, A. Forchel, and Y. Yamamoto, Nature Physics \textbf{4}, 700 (2008).

\bibitem{Ciuti1998} C. Ciuti, V. Savona, C. Piermarocchi, and A. Quattropani, P. Schwendimann, Phys. Rev. B \textbf{58}, 7926 (1998).

\bibitem{Combescot2007} M. Combescot, O. Betbeder-Matibet, and R. Combescot, Phys. Rev. Lett. \textbf{99}, 176403 (2007).

\bibitem{Renucci} P. Renucci, T. Amand, X. Marie, P. Senellart, J. Bloch, B. Sermage, and K. V. Kavokin, Phys. Rev. B \textbf{72}, 075317 (2005).

\bibitem{Laussy2006} F.P. Laussy, I.A. Shelykh, G. Malpuech, and A. Kavokin, Phys. Rev. B \textbf{73}, 035315 (2006).

\bibitem{Shelykh2006}  I. A. Shelykh, Yuri G. Rubo, G. Malpuech, D. D. Solnyshkov, and A. Kavokin, Phys. Rev. Lett. \textbf{97}, 066402 (2006).

\bibitem{PolDevices} T. Liew, I.A. Shelykh, G. Malpuech, Review: Polaritonic Devices, to appear in Physica E.

\bibitem{Zagoskin} A. M. Zagoskin, \textit{Quantum Theory of Many-Body Systems} (Springer, New York, 1998).

\bibitem{Ouerdane} M. M. Glazov, H. Ouerdane, L. Pilozzi, G. Malpuech, A. V. Kavokin, and A. D'Andrea Phys. Rev. B \textbf{80}, 155306 (2009).

\bibitem{Ostatnicky} T. Ostatnicky, D. Read, and A.V. Kavokin, Phys. Rev. B \textbf{80}, 115328 (2009).

\bibitem{Masha} M. Vladimirova, S. Cronenberger, D. Scalbert, K. V. Kavokin, A. Miard, A. Lemaitre, J. Bloch, D. Solnyshkov, G. Malpuech, and A. V. Kavokin, Phys. Rev. B \textbf{82}, 075301 (2010).

\bibitem{Tassone} F. Tassone and Y. Yamamoto, Phys. Rev. B \textbf{59}, 10830 (1999).

\bibitem{Lindhard} J. Lindhard, Dan. Math. Phys. Medd. \textbf{28}, 8 (1954).

\bibitem{Koch} H. Haug, S. W. Koch, \textit{Quantum Theory of the Optical and Electronic Properties of Semiconductors} (World Scientific, Singapore, 1990).

\bibitem{Stern} F. Stern, Phys. Rev. Lett. \textbf{18}, 14, 546-548 (1967).

\bibitem{Ramon2002} G. Ramon, R. Rapaport, A. Qarry, E. Cohen, A. Mann, Arza Ron, and L. N. Pfeiffer, Phys. Rev. B \textbf{65}, 085323 (2002).

\bibitem{Ramon2003} G. Ramon, A. Mann, and E. Cohen, Phys. Rev. B \textbf{67}, 045323 (2003).

\bibitem{Tamm} Compare to anticrossing between Tamm plasmon and polariton modes in hybrid cavity- metal structures, M. Kaliteevski, S. Brand, R. A. Abram, I. Iorsh, A. V. Kavokin, and I. A. Shelykh, Appl. Phys. Lett. \textbf{95}, 251108 (2009).

\bibitem{RuboPLA} Yu. G. Rubo, A.V. Kavokin, I.A. Shelykh, Phys. Lett. A \textbf{358}, 227 (2006).

\bibitem{Christmann} G. Christmann, C. Coulson, J. J. Baumberg, N.T. Pelekanos, Z. Hatzopoulos, S.I. Tsintzos, and P.G. Savvidis, Phys. Rev. B \textbf{82}, 113308 (2010).

\bibitem{Oriented} G. Christmann, J. J. Baumberg, A. Askitopoulos, G. Deligeorgis, Z. Hatzopoulos, S. I. Tsintzos, and P. G. Savvidis,  arXiv:1102.2852, to appear in Appl. Phys. Lett.

\end{thebibliography}
\end{document}